\documentstyle[12pt,epsf]{article}

\begin{document}

\newcommand{\myprime}{^\prime}
\newcommand{\grad}{\nabla}

\title{\begin{flushright}
\begin{small} 
hep-th/9605174,  PUPT-1600 
\end{small} 
\end{flushright}
Some Aspects of Massive World-Brane Dynamics}
\author{Vijay Balasubramanian 
and Igor R. Klebanov\thanks{vijayb,klebanov@puhep1.princeton.edu}\\
       {\it Joseph Henry Laboratories,} \\
       {\it     Princeton University,
         Princeton,   NJ 08544}}
\date{\today}                                 
\maketitle                                 

\begin{abstract}
We study the internal dynamics of Ramond-Ramond solitons excited far
from the BPS limit by leading Regge trajectory open strings.
The simplest world volume process for such strings is splitting into two
smaller pieces, and we calculate the corresponding decay rates.
Compared to the conventional open superstring, the
splitting of states polarized parallel to the brane is suppressed by
powers of logarithms of the energy.  The rate for states polarized
transverse to the brane {\em decreases} with increasing energy.  We
also calculate the static force between
a D-brane excited by a massive open string and an unexcited
D-brane parallel to it. The result shows that transversely
polarized massive open strings endow D-branes with a size of order
the string scale.
\end{abstract}

\section{Introduction}
\label{sec:intro}
Open strings with Dirichlet boundary conditions provide a conformal
field theory description of Ramond-Ramond solitons~\cite{polch95a}.
In this paper we study the dynamics of these objects moved far from
the BPS limit by states on the leading Regge trajectory.  We are
motivated in this study by recent progress in describing black hole
thermodynamics using D-branes~
\cite{strom96a} -~\cite{vijay96b}.
In order to have a thermodynamic description, excitations of an object
must be able to equipartition their energy at a sufficiently rapid
rate.  The D-brane black holes  constructed thus far involve
multiple intersecting branes and are difficult to handle in such a
dynamical context.  On the other hand, individual p-branes have
classical horizons~\cite{horstrom} and in some cases have entropies
that scale in the same way as the Bekenstein-Hawking
formula~\cite{gubser,igor96b}.  Therefore, we hope to gain some insight into
generic issues like equipartition of energy by studying individual RR
solitons. 

To this end, we excite D-branes far from the BPS limit by attaching
massive open strings on the leading Regge trajectory to them. 
The importance of massive excitations on the world volume is
exhibited by scattering of closed strings off D-branes
\cite{kthor,ghkm,myers}. There, the massive intermediate states
give rise to an infinite sequence of $s$-channel poles. It is
precisely these excitations that endow D-branes with an effective
transverse size which increases with the energy of the probe and is of
order $\sqrt{\alpha'}$. The dynamics of the massive modes is not constrained
by supersymmetry, and in this paper we explore some of its features at
weak coupling. We find it important to distinguish between states that
are polarized in the longitudinal and in the transverse
directions. The former classically correspond to strings that lie
entirely within the brane, while the latter describe strings
``hanging'' off the brane. We will find that there are major
differences in the behavior of these two kinds of states. Not
surprisingly, it is only the transversely polarized states that are
responsible for the energy-dependent thickness of D-branes (we
demonstrate this is Section 4).

The dominant world volume
process for the massive
strings is splitting into pairs of lighter states.
The rate for the splitting process can be studied by extracting the
imaginary part of the one-loop diagram for strings attached to a
D-brane.  We find that the decay rates are suppressed compared to the
results for a conventional superstring.  States polarized parallel to
the brane have decay rates suppressed by powers of logarithms of the
energy.  States polarized transverse to the brane have rates that
often {\em decrease} with increasing energy.  Furthermore, strings
attached to a 0-brane simply cannot decay via splitting into two
smaller pieces - the leading channels are higher order in the string
coupling.  The source of these unusual effects is the fact that all
open string momenta must lie within the world volume.

In Section~\ref{sec:amp} we calculate the one-loop amplitude for a
D-brane.  We employ it in Section~\ref{sec:rates} to extract the decay
rates of highly excited string states.  In Section~\ref{sec:forces}
the one-loop amplitude is used to study forces between parallel
D-branes.  Evidence is displayed that excitations transverse to a
brane give it a finite extent which increases with the excitation
level and is of order the string length.  Implications of the results
are discussed in Section~\ref{sec:impl}.

\section{D-Brane Loop Amplitude}
\label{sec:amp}
\begin{figure}                                 
\begin{center}                                 
\leavevmode                                 
\epsfxsize=1in
\epsfysize=1in                                 
\epsffile{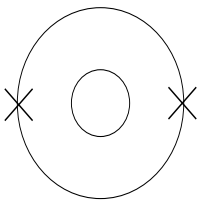}                                 
\end{center}                                 
\caption{When both boundaries are placed on the same brane, the
imaginary part of this diagram gives the decay rate of a state via
splitting into pairs of lighter states attached to the brane.  When
the boundaries are on different branes the diagram computes the static
interaction potential between an excited brane and another in its
ground state. \label{fig:one}}
\end{figure}

The decay rates we are interested in are most easily calculated as the
imaginary part of the annulus diagram in
Figure~\ref{fig:one}.\footnote{Our strings are oriented, so the
M\"{o}bius strip is absent.  The nonplanar contribution is expected to
be highly suppressed, and contains closed string intermediate
states~\cite{mtwj,mst,kls}.}  As an added benefit, when the boundaries
of the annulus are attached to
different parallel branes, the diagram gives the
interaction potential between them.  We will calculate the annulus
amplitude for NS states on the
leading Regge trajectory by modifying the
results of~\cite{yamamoto}, obtained for the conventional open superstring.

It is useful to review the analysis of the one-loop two-point
amplitude of the open superstring.  On the annulus we can choose
vertex operators of ghost number zero~\cite{kls}.  For leading Regge
trajectory states with $m^2 = 2l$ and $J = l+1$ for $l \geq 0$ this
gives~\cite{yamamoto}:
\begin{eqnarray}
V_{(2l,l+1)} =
\frac{\xi_{\mu\nu_1\cdots\nu_l}}{\sqrt{l!}} \left[
\partial X^\mu \partial X^{\nu_1}\cdots \partial X^{\nu_l}
+ \sum_{j=1}^l \partial\psi^{\nu_j} \, \psi^\mu \,
\partial X^{\nu_1} \cdots
\partial X^{\nu_{j-1}} \partial X^{\nu_{j+1}} \cdots
\partial X^{\nu_l} \right. \nonumber \\
+ \left. i (k\cdot \psi) \, \psi^\mu \,
\partial X^{\nu_1} \cdots \partial X^{\nu_l}
\right] \: e^{i k \cdot X}
\label{eq:vertex}
\end{eqnarray}
Here $\partial$ stands for $\partial_\tau$, the derivative with
respect to the world-sheet time and $\xi$ is a totally symmetric
polarization tensor that is transverse to the momentum $k$ and traceless.

The two-point annulus amplitude for these vertex operators has the
general structure: 
\begin{equation}
A = \sum_s \eta_s \int d\omega  \: F_s(\omega) \: \langle V_1 V_2
\rangle_s  \label{eq:general}
\end{equation}
where $\sum_s$ is the sum over spin structures, $\eta_s$ is relative
phase between spin structures and $d\omega$ is a measure on the
annulus.  The factors $F_s$ include the trace over the momentum
running around the loop and the partition function of oscillators, and
the correlation functions are evaluated separately in each spin
structure.  It is convenient to work in variables $w = \rho_1 \rho_2$
and $\nu = \ln{\rho_1}/\ln{w}$ where $\rho_1$ and $\rho_2$ are the
standard variables on the strip introduced in Chapter 8 of~\cite{gsw}.
Then $\ln{w}$ is the world-sheet time elapsed in travelling around the
loop.  Evaluating the correlation functions and doing the sum over
spin structures gives~\cite{yamamoto}:
\begin{eqnarray}
A &=& {\cal N} \int_0^1 d\nu \int_\epsilon^\infty d(-\ln{w})
\left(\frac{1}{-\ln{w}}\right)^4 \: \Psi(\nu, w)^{2l}  \:
\Omega(\nu,w)^{l-1} \nonumber \\
{\cal N} &=&  2\pi^2 \, (2\pi)^3 \,\xi^2 \, l \label{eq:amp1}
\end{eqnarray}
where $\Psi = \langle X X \rangle$ and $\Omega = \langle \partial X
\partial X \rangle$ are the bosonic correlators; 
$\xi^2$ stands for 
$\xi^{(1)}_{\mu\nu_1\cdots\nu_l}\xi^{(2)}_{\mu\nu_1\cdots\nu_l}$.
As usual, we have
introduced $\epsilon$, a cutoff that is necessary to define the
integral over the world-sheet momenta and is taken to zero at the end
of the calculation. The integration over these momenta gives
rise to a factor $(-2\pi/\ln{w})^{D/2} = (-2\pi/\ln{w})^5$ in
Eq.~\ref{eq:amp1}.\footnote{The additional factor of $(-\ln{w})$ in
Eq.~\ref{eq:amp1} arises from the measure for $d\nu$~\cite{gsw}.}
Note that the physical meaning of the annulus amplitude is the
one-loop mass shift. The fact that this vanishes for the gauge boson
($l=0$ state) is in accord with gauge invariance. 

In order to generalize to D-brane amplitudes, note that the partition
functions and correlators appearing in Eq.~\ref{eq:general} factorize
into products of fermionic and bosonic parts.  We will first study the
effect of Dirichlet boundary conditions on bosons and then on
fermions.  Dirichlet boundary conditions are implemented on some
coordinates by flipping the sign of the corresponding left-moving
bosons.  This yields the following mode expansion describing strings
stretched between parallel k-branes.
\begin{eqnarray}
0\leq\mu\leq k :  \: \: \: \: \:      
X^\mu &=& X_0^\mu + p^\mu \tau + 
i \, \sum_{n \neq 0} \frac{1}{n} \alpha_n^\mu \cos(n\sigma) e^{-in\tau} 
\nonumber \\
k+1 \leq \mu \leq D :  \: \: \: \: \:
X^\mu &=& X_0^\mu + \frac{Y^\mu}{\pi} \sigma + 
\sum_{n \neq 0} \frac{1}{n} \alpha_n^\mu \sin(n\sigma) e^{-in\tau} 
\label{eq:bmodes}
\end{eqnarray}
Since the branes are stationary, we do not integrate over $Y$, which
measures the offset between them.  Furthermore, neither the vertex
operators on the boundary of Figure~\ref{fig:one} nor the states
running around the loop can carry momentum in the Dirichlet
directions.  Finally, the change in sign of the left moving boson
implies that the operators $\partial X^\mu$ appearing in
Equation~\ref{eq:vertex} should be read as $\partial_\tau X^\mu$
(tangential derivatives) for Neumann coordinates and $\partial_\sigma
X^\mu$ (normal derivatives) for Dirichlet coordinates.

Since the Dirichlet boundary conditions do not affect the spectrum of
the open strings, the bosonic contribution to the partition functions
in Eq.~\ref{eq:general} is unchanged.  However, the restriction of
world-sheet momenta to lie parallel to the brane has important
effects.  First of all, the trace over world-sheet momenta produces a
factor of $(-2\pi/\ln{w})^{(k+1)/2)}$ instead of
$(-2\pi/\ln{w})^{D/2}$.  Furthermore, the separation $Y^\mu$ between
the boundaries contributes a factor of $\exp{(Y^2
\ln{w}/2\pi^2)}$.\footnote{The separation between the branes
contributes $Y^2/(4\pi^2\alpha^{\prime})$ to $L_0$~\cite{polch96a}.
Setting $\alpha^\prime = 1/2$, this adds a contribution
$w^{Y^2/(2\pi^2)}$ to the loop amplitude.}  Finally, the bosonic
correlators are modified as discussed in~\cite{gutperle,cmnp}.
Further subtleties arise when the separation $Y^\mu$ is not orthogonal
to the polarization tensor.  We will avoid them by only considering
states polarized transverse to both the momentum $k$ and the
separation $Y$.  In any case, we are mostly interested in the case $Y
= 0$ for which the additional transversality requirement does not pose
a constraint.


On the other hand, the fermionic contributions to the amplitude are
completely unchanged by the imposition of Dirichlet boundary
condition.  To see this, remember that Dirichlet boundary conditions are
implemented by changing the sign of the left moving bosons, and so
supersymmetry instructs us to flip the sign of the left-moving
fermions also.  Now, the NS and R sectors of the amplitude arise from the
relative sign of the left and right moving fermions at the ends of the
string.  But changing the sign of the left-mover only changes the
conventional overall sign between left-moving and right-moving
fermions.  Therefore imposing Dirichlet boundary conditions on some
coordinates has no effect on the fermionic partition function and
correlators.

Using the factorization of the bosonic and fermionic contributions to
the one-loop amplitude and the observations made above, the results
of~\cite{yamamoto} are readily combined with the correlators derived
in~\cite{gutperle} to yield D-brane amplitudes.  It is easiest to work
with states polarized entirely parallel or perpendicular to a
brane. (As discussed above, the states are also taken to be
orthogonal to the separation $Y^\mu$.)  The one-loop amplitudes for
leading Regge trajectory states at level $l$ on a k-brane are:
\begin{eqnarray}
A_{(N,D)} &=& {\cal N}_{(N,D)} \int_0^1 d\nu \,
\int_\epsilon^\infty d(-\ln{w})
\left(\frac{1}{-\ln{w}}\right)^{\frac{k-1}{2} }
e^{(Y^2/2\pi^2)\ln{w}} \:
\Psi^{2l} \: \Omega_{(N,D)}^{l-1} \label{eq:amp}\\ 
\Psi &=&  (-\ln{w}) \: \frac{\theta_1(\nu|\tau)}{\theta_1^\prime(0|\tau)}
= \frac{e^{(1/2) \nu (\nu -1 ) \ln{w}}}{f(w)^3}
\sum_{n=-\infty}^\infty (-1)^n \rho_1^{n(n+1)/2} \rho_2^{n(n-1)/2}
 \label{eq:psi}\\
\Omega_N &=& \frac{-1}{(\ln{w})^2} \: \partial_\nu^2
\ln{\theta_1(\nu|\tau)} =
\frac{-1}{\ln{w}} + 
\sum_{n=1}^\infty n \left[ \frac{\rho_1^n + \rho_2^n}{1 - w^n} \right]
\label{eq:omegan} \\
\Omega_D &=& \Omega_N + \frac{1}{\ln{w}} = 
\sum_{n=1}^\infty n \left[ \frac{\rho_1^n + \rho_2^n}{1 - w^n} \right]
 \label{eq:omegad} \\
{\cal N}_{(N,D)} & =&  (2 \pi^2) \, (2\pi)^3 \:  
(2\pi)^{(k+1 -D)/2} \:
l \:
\xi_{(N,D)}^2 
 \: \: \: \: \: \: \: \: ;
 \: \: \: \: \: \: \: \: 
f(w) = \prod_{i=1}^\infty (1 - w^i) 
 \label{eq:coeff}
\end{eqnarray}
Here $\rho_1 = \exp{(\nu \ln{w})}$ and $\rho_2 = \exp{((1 - \nu)
\ln{w})}$ with $w = \rho_1 \, \rho_2$ and $\tau = -2\pi i/\ln{w}$.  We
have used standard series expansions for the $\theta$ functions
arising in the $\Psi$ and $\Omega$ correlators.  The subscripts N and
D refer to states that are polarized
entirely in the longitudinal (Neumann) or
transverse (Dirichlet) directions respectively.  The difference of
a $-1/\ln{w}$ between $\Omega_N$ and $\Omega_D$ will have a profound
effect on the relative decay rates.

\section{Splitting Rates}
\label{sec:rates}
The splitting rates of highly excited string states can be extracted from
Eq.~\ref{eq:amp}.  In the old operator formalism, the overall
amplitude is written as $Tr(\rho_1^{L_0} V_1 \rho_2^{L_0} V_2)$.  So,
the term proportional to $\rho_1^p \rho_2^q$ in a power series
expansion of the integrand contains the intermediate channel of levels
$p$ and $q$.\footnote{See~\cite{mtwj} and~\cite{okada} for similar
analyses of conventional open strings.}  Performing such an expansion
to isolate the channels gives:
\begin{eqnarray}
A_N &=& {\cal N}_N \sum_{p,q \geq 0} \: \sum_{r=0}^{l-1} \, 
C_{pq}^{r} \, \int_0^1 d\nu \int_\epsilon^\infty d(-\ln{w})
\, \left(\frac{1}{-\ln{w}}\right)^{r + (k-1)/2 } \, 
e^{B(\nu, p, q) \ln{w}}
\label{eq:an} \\
A_D &=& {\cal N}_D \sum_{p,q \geq 0} \, 
D_{pq}  \, \int_0^1 d\nu \int_\epsilon^\infty d(-\ln{w})
\, \left(\frac{1}{-\ln{w}}\right)^{(k-1)/2 } \, 
e^{B(\nu, p, q) \ln{w}}
\label{eq:ad} \\
B(\nu,p,q) &=& l(\nu-1)\nu + p \nu + q ( 1 - \nu) + \frac{Y^2}{2\pi^2}
\label{eq:A} 
\end{eqnarray}
Here the index $r$ is associated with powers of $\ln(w)$ that arise
from expansion of $\Omega_N^{l-1}$ in the case of parallel
polarizations.  The coefficients $C_{pq}^r$ and $D_{pq}$ are
determined by expanding the integrand of Eq.~\ref{eq:amp} in powers of
$\rho_1$ and $\rho_2$.  When $B < 0$ this (Euclidean) amplitude
diverges, and it is easily seen that this can only happen when
$\sqrt{l} >\sqrt{ p + Y^2/4\pi^2} + \sqrt{q + Y^2/4\pi^2}$ as should
be expected for a threshold cut.\footnote{In fact, because we are
working with a Euclidean world-sheet, the amplitudes are formally
real.  However, we will shortly continue back to a Minkowski
world-sheet in order to extract the decay amplitudes.}  Below, we set
$Y^2 = 0$ and study the imaginary part of the amplitude arising from
these threshold cuts to derive the decay rate of highly excited string
states on D-branes.

\subsection{Transverse Polarizations}
\label{sec:trans}

\begin{figure}                                 
\begin{center}                                 
\leavevmode                                 
\epsfxsize=1in
\epsfysize=1in                                 
\epsffile{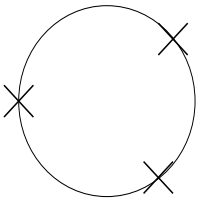}                                 
\end{center}                                 
\caption{The leading channel for splitting of strings attached to a
brane is this three point diagram. \label{fig:three}}
\end{figure}

For decays of traversely polarized states the key point is that the
absence of a leading $(-1/\ln{w})$ in $\Omega_D$ implies that  $p + q
\geq l-1$ in $A_D$.   The threshold condition for the $(p,q)$ channel
requires $\sqrt{l} > \sqrt{p} + \sqrt{q}$ so that $l > p + q +
2\sqrt{pq}$.   The only simultaneous solutions of these equations are
$(p,q) = \{ (l-1,0), (0,l-1) \}$.    
Thus, the only decay channel for a transversely
polarized leading Regge trajectory
state at level $l$ is into a leading Regge trajectory
state at level $l-1$ and a massless state.

This can also be seen from the disk diagram in Figure~\ref{fig:three}
representing a given decay channel.  Suppose one of the boundary
operators is a level $l$, transversely polarised, leading Regge
trajectory state and the other two are states at levels $p$ and $q$.
Then, by conservation of energy, $\sqrt{l} \geq \sqrt{p} + \sqrt{q}$.
Next, in the center of mass frame, the first vertex operator carries
an ``angular momentum'' $l+1$\footnote{More precisely, this means that
the bosonic part of this vertex operator contains $l+1$
$\partial_n X$ factors.} which must be conserved by the decay
products.  The angular momentum carried by a stationary transversely
polarized state cannot be made up from the relative motions of the
decay products on the brane.  So, because the intrinsic angular
momentum carried by a state at level $p$ is at most $p+1$, we must
have $ (p + 1) + (q + 1) \geq l +1$ giving $p + q \geq l-1$.  This
reproduces the selection rules for decays of transversely polarized
states derived in the previous paragraph.

\subsection{$k>0$}

\begin{figure}                                 
\begin{center}                                 
\leavevmode                                 
\epsfxsize=5in
\epsfysize=1in                                 
\epsffile{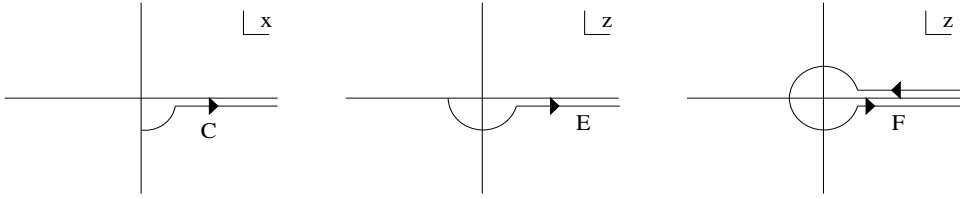}                                 
\end{center}                                 
\caption{Integration contours for decay amplitudes. \label{fig:contours}}
\end{figure}

For a k-brane with $k>0$ the decay amplitude is most readily extracted
by continuing Eq.~\ref{eq:ad} back to a Minkowski world-sheet and
examining the imaginary part.   Setting $ -\ln{w} = ix$, the integral
in the $(p,q)$ channel is:
\begin{equation}
I_{pq} = \int d\nu \int_C dx \,i\, 
(ix)^{(1 - k)/2} e^{-i x B(\nu,p,q)}
\label{eq:int1}
\end{equation}
where C is the contour in Figure~\ref{fig:contours}.\footnote{The
integral starts on the negative imaginary axis in order to make the
trace over world-sheet Minkowski momenta well defined.  The offset is
taken to zero at the end of the calculation. See~\cite{mtwj} for a
similar treatment of the bosonic string.}  The integral can be carried
out by rotating the contour to the positive or negative imaginary
axes. When $B>0$, the contour must be rotated down, giving a purely
real integral.  When $B<0$, the contour must be rotated up and picks
up a contribution from the pole at the origin.\footnote{
For  $k=1$ there is apparently no divergence in Eq.~\ref{eq:ad} at $\ln{w}
\approx 0$.   However, the case of a 1-brane is rather delicate since
because the imaginary part of the amplitude for $k = 1 + \epsilon$ is
non-zero while it vanishes for $k= 1 - \epsilon$.  The physical
prescription is to define the decay amplitude for a 1-brane by a
continuation from $k=1 + \epsilon$ as will be seen below from the
agreement of the computed decay rate with phase space arguments.}
  Letting $\nu_+$ and
$\nu_-$ be the two zeroes of $B(\nu,p,q)$, and performing some
rescalings, the imaginary part of the amplitude in the $(p,q)$ channel
is:
\begin{eqnarray}
Im(A_{pq}) &=& {\cal N}_D \, D_{pq} \int_{\nu_-}^{\nu_+}  d\nu \,
                  |B(\nu,p,q)|^{(k-3)/2} Im(J) \label{eq:int2}  \\
Im(J) &=& - Im\left[ \int_E dz \, (-z)^{(1 - k)/2} \, e^{-z} \right]
\nonumber \\
 &=& \frac{-1}{2} Im\left[ \int_F dz  \, (-z)^{(1 - k)/2} \, e^{-z} \right]
 = \frac{\pi}{\Gamma((k-1)/2)} \label{eq:int3}
\end{eqnarray}
The contours E and F are shown in Figure~\ref{fig:contours} and the
last equality in Eq.~\ref{eq:int3} follows from H\"{a}nkel's formula for the
Gamma function~\cite{ww}.   Defining $a = (\nu_+ - \nu_-)/2$, the
remaining integral can be done as in~\cite{mtwj} to find the decay
rate into the $(p,q)$ channel:
\begin{equation}
Im(A_{pq}) = {\cal N}_D \, D_{pq} \,  
\frac{\pi^{3/2} \, l^{(k-3)/2} \, a^{k -2}}{\Gamma(k/2)}
\end{equation}
Finally, $a = 1/(2l)$
for the channels $(l-1,0)$ and $(0,l-1)$.  Combining this with
$D_{l-1,0}= 1$ in Eq.~\ref{eq:ad} and the relativistic normalization
$\xi_D^2 = 1/2\sqrt{l}$~\cite{mtwj}, yields the decay rate:
\begin{equation}
Im(A_D) = \left(\frac{1}{2\pi}\right)^4 \:
\frac{ 8\pi^5\,\pi^{3/2}}{\Gamma(k/2)} \: \sqrt{l} \:
\left(\frac{\pi}{2l}\right)^{(k-1)/2}
\label{eq:rated}
\end{equation}
Surprisingly, for $k>2$ the decay rate actually {\em decreases} as the
level $l$ increases, while for $k=2$ the rate is independent of
$l$. Naively one might have expected that the decay rate grows with
the excitation number.  This does not happen due to the extremely
constrained phase space for the decays of transversely polarized
states.  Indeed, in the center of mass frame, energy conservation for
the decay of a level $l$ state into states of levels $l-1$ and $0$
requires the decay products to carry spatial momentum with magnitude
of order $1/\sqrt{l}$.  The available phase space for states with this
momentum scales as $(1/l)^{(k-1)/2}$ and explains the final
suppression factor in Eq.~\ref{eq:rated}.  In particular, note that on
a 1-brane energy conservation gives a decay phase space consisting of
precisely two points.  The volume of phase space is therefore
independent of energy.  This agrees with the lack of suppression in
Eq.~\ref{eq:rated} for $k=1$.

\subsubsection{$k=0$}
\label{sec:k01}
When $k=0$, the absence of a divergence in Eq.~\ref{eq:ad}  at $\ln(w)
\approx 0$ tells us that the procedure of the previous section will
not give us an imaginary part.  So the rate for splitting
of a level $l$ state on a 0-brane into two smaller pieces is exactly
zero.   Indeed, for a 0-brane, conservation of energy requires that
$\sqrt l= \sqrt p+\sqrt q$, which is in immediate
conflict with the selection rule $\sqrt{l} >
\sqrt{p} + \sqrt{q}$.   
Thus, there are no allowed decay channels. This statement also applies
to subleading Regge trajectories since it follows from energy
conservation on a $0+1$ dimensional world-volume.  Therefore, excited
states of 0-branes are exceptionally stable - they decay via higher
order diagrams involving more powers of the string coupling.  The
physical implication of this fact is the relative stability of 
excitations of ten-dimensional RR-charged black holes - they can only
decay via emission of closed string states.


\subsection{Parallel Polarizations}
\label{sec:para}
Because the expansion of $\Omega_N^{l-1}$ in powers of $\rho_1^p$ and
$\rho_2^q$ contains all terms with $p+q \leq l-1$, all channels that
are permitted by the threshold condition contribute to the decays.
The combinatorics of the channel-by-channel sum is very hard and so we
instead carry out an asymptotic analysis of the total decay
rate~\cite{mst}.   

The key point  is that the imaginary part of the
one-loop amplitude arises from the nature of the divergence near $w
= 1$.  The analysis is easiest in annular variables $\nu$ and $\ln{q} =
2\pi^2/\ln{w}$.    Transforming to these variables, and using the modular
properties of $\theta$ functions yields the amplitude:
\begin{equation}
A_N = \frac{{\cal N}_N}{2\pi^2}\, 
\int_0^1 d\nu \int_0^1 \frac{dq}{q} 
\left(\frac{-\ln{q}}{2\pi^2}\right)^{(k+1-D)/2}
e^{Y^2/\ln(q)}
\left( \frac{\theta_1(\nu|\tau)}{\theta_1\myprime(0|\tau)}
\right)^{2l}
(-\partial_\nu^2\ln{\theta_1(\nu|\tau)})^{l-1}
\label{eq:nampq}\\
\end{equation}
We now set $Y^2 = 0 $ because we are interested in decays of states
attached to a single brane.  The behaviour of the amplitude near $q=0$
(the ultraviolet region of the open string channel) determines the
decay rate.  Expanding the amplitude near $q=0$ using standard
$\theta$ function identities gives:
\begin{equation}
A_{N} = \frac{{\cal N}_N}{2\pi^2} \int_0^1 \frac{dq}{q} \int_0^1 d\nu 
\left(\frac{\sin^2(\pi\nu)}{\pi^2}\right) 
\left(\frac{-\ln(q)}{2\pi^2}\right)^{(k+1-D)/2}
\exp\left[ 16 l q^2 \sin^4(\pi\nu) \right] \label{eq:smallqn}
\end{equation}

To extract the splitting rate from this amplitude we follow
the work of~\cite{mst} and change variables to $z = \sin^2(\pi\nu)$
and $y = q^2 z^2$.  We are not interested in the singularities of the
amplitude that arise when the two operators on the boundary coincide.
So the entire analysis implicitly contains a cutoff
that bounds $z$ away from zero and is removed after the rest of
computations are carried out.   Keeping this cutoff in mind so that
$\ln(q^2z^2) \approx 2\ln(q)$ for small $q$, the  amplitude in the new
variables becomes: 
\begin{equation}
A_{N} = \frac{{\cal N}_N}{2\pi^3(2\pi^2)} \int_0^1 dz \frac{z}{\sqrt{z}\sqrt{1-z}}
\int_0^{z^2} \frac{dy}{y}
\left(\frac{-\ln(y)}{4\pi^2}\right)^{{\textstyle{\frac{k+1-D}{2}}}} 
e^{{\textstyle 16ly}}
\end{equation}
The inner integral is only well-approximated by small values of $q$ if
$l$ is large and negative.  So we analytically continue the amplitude
in the complex $l$ plane by setting $l \rightarrow - \bar{l}$.  The
$y$ integral can now be reliably estimated by approximating
$\exp{-16\bar{l} y}$ as $1$ for $0\leq y \leq 1/(16\bar{l})$ and $0$
for $y > 1/(16\bar{l})$.   The estimate of the analytically continued
$y$ integral is:
\begin{equation}
E = \frac{-2}{(k+3-D)}
\left(\frac{1}{4\pi^2}\right)^{{\textstyle \frac{k+1-D}{2}}}
\left(\ln{16\bar{l}}\right)^{{\textstyle \frac{k+3 - D}{2}}}
\end{equation}
The integral over $z$ can now be done independently and gives
$B(3/2,1/2)=\pi/2$.   Multiplying the factors
together, rotating $l$ back to positive real line ($\bar{l} \rightarrow
- \bar{l}$ ) and taking the leading imaginary part of the amplitude gives:
\begin{equation}
Im(A_N) = 2\pi^2\,
\sqrt{l} \,
\left( \frac{\ln(l)}{2\pi} \right)^{{\textstyle \frac{k+1-D}{2}}}
\end{equation}
where we used the normalization $\xi_N^2 = 1/2\sqrt{l}$.  The
splitting rate is given by $Im(A_N)$.  We recover the
result for a conventional open string by taking $k=9$.  In that case,
the decay rate is proportional to $\sqrt{l}$ as it should be for a
splitting rate per unit length.  For strings attached to a general
k-brane the decay rate is suppressed by logarithms of the energy.
These logarithms arose from the restriction to the Neumann directions
of the momenta running around the loop.  Equivalently, they arose from
a truncation of intermediate open string states travelling normal to
brane and the consequent restriction of the decay phase space.

An asymptotic analysis of this type would not have worked for the
transversely polarized states.  In that case the important region of
the $\nu$ integral was in the neighbourhood of 0 and 1 which is
precisely the region cut off in this analysis.  Essentially, the
dominant decays of states polarized parallel to the brane are into
channels where both decay products are at low levels compared to the
original state.  These decays are allowed because angular momentum can
be transferred to the relative spatial motion of the decay products
and have a relatively large phase space.  This is in marked contrast to
the situation for transversely polarized states.

\section{Forces Between Branes}
\label{sec:forces}
The one-loop amplitude in Eq.~\ref{eq:amp} with nonzero $Y$ can be
interpreted as the static interaction potential between a D-brane
carrying massive excitations and a parallel D-brane in its ground state a
distance $Y^\mu$ away.  This interpretation follows by thinking of the
annulus diagram in the cross-channel as a closed string exchanged by
D-branes in position eigenstates.  To study long range forces we want to
extract the contribution of the massless closed string states which
arises from the $w \rightarrow 1$ limit of Eq.~\ref{eq:amp}. This limit
is most conveniently extracted as the $q \rightarrow 0$ limit in the
annular variable $\ln{q} = 2\pi^2/\ln{w}$ introduced in
Section~\ref{sec:para}.

\paragraph{Parallel Polarizations:}  The $q \rightarrow 0$ limit  has
already been extracted for parallel polarizations in Eq.~\ref{eq:smallqn},
except that we must restore the factor of $\exp(Y^2/ \ln{q})$ that
was dropped there.  For small $q$ this term dominates over the
exponential of $q^2$ that appeared in Eq.~\ref{eq:smallqn} so we can
drop the latter term in computing long-range forces.  Changing
variables to $z = \sin^2(\pi \nu)$ and $t = -1/\ln(q)$ we find that the
contribution of the closed string massless states is:
\begin{equation}
A_{N} = \frac{{\cal N}_N}{2\pi^2\pi^3} {\pi^3} \int_0^1 \frac{dz \,
z}{\sqrt{z}\sqrt{1-z}}
\int_0^\infty \frac{dt}{t^2} 
\left(\frac{1}{2\pi^2 t}\right)^{{\textstyle \frac{k+1-D}{2}}}
e^{-tY^2}
\end{equation}
Doing the integrals and putting in the  normalizing factor ${\cal N}$
gives:
\begin{equation}
A_{N} = 
2\pi \,\sqrt{l} \, 
\left(\frac{1}{\pi}\right)^{{\textstyle \frac{k+1-D}{2}}}
\Gamma\left(\frac{D-3-k}{2}\right)
\left(\frac{1}{Y^2}\right)^{{\textstyle \frac{D - 3 - k}{2}}}
\end{equation}
$A_{N}$ is proportional to the mass of the excitation $\sqrt{l}$ times
the $9-k$ dimensional Green's function $G_{9-k}(Y^2)$.  Exciting
the open string states on the first brane has broken the residual
supersymmetry of the BPS state and this is reflected in a
non-vanishing force.  We see a long-range Coulombic force
acting between the parallel branes that is proportional to $\sqrt{l}$,
the mass excess over the BPS bound.  The force vanishes for $l = 0$
because parallel massless excitations of a brane do not destroy the
BPS saturation. This provides a nice check on our result.

\paragraph{Transverse Polarizations:}  Repeating the analysis of
small $q$ asymptotics for transverse polarizations gives:\footnote{To
arrive at this expression, we use standard $\theta$ function
identities to find that in the small $q$ limit the leading term in the
integrand of Eq.~\ref{eq:amp} contains a factor $(1 + 2
\sin^2(\pi\nu)/\ln{q})^l$.  We write this as $\exp\left[l \, \ln(1 + 2
\sin^2(\pi\nu)/\ln{q})\right]$ and keep the leading term in the
exponent.  This treatment is valid for large $l$.}
\begin{equation}
A_{D} = \frac{{\cal N}_D}{2\pi^2}  \int_0^1 d\nu \int_0^1 \frac{dq}{q}
\left(\frac{\sin^2(\pi\nu)}{\pi^2}\right) 
\left(\frac{-\ln(q)}{2\pi^2}\right)^{(k+1-D)/2}
\exp\left[ (Y^2 + 2 l \sin^2(\pi\nu))/\ln(q) \right]
\label{eq:smallqd}
\end{equation}
Changing variables to $z$ and $t$ as above gives:
\begin{equation}
A_{D} = \frac{{\cal N}_D}{2\pi^2 \pi^3} \int_0^1 \frac{dz \,
z}{\sqrt{z}\,\sqrt{1-z}}
\int_0^\infty \frac{dt}{t^2} 
\left(\frac{1}{2\pi^2 t}\right)^{{\textstyle \frac{k+1-D}{2}}}
e^{-t(Y^2 + 2lz)}
\label{eq:dnewvar}
\end{equation}
Doing the integrals and inserting ${\cal N}_D$ gives a static
interaction potential of:
\begin{eqnarray}
A_{D} &=& 4\,\sqrt{l}\, 
\left(\frac{1}{\pi}\right)^{{\textstyle \frac{k+1-D}{2}}}
\Gamma\left(\frac{D-3-k}{2}\right)
\int_0^1 \frac{dz \, z}{\sqrt{z} \, \sqrt{1-z}} 
\left(\frac{1}{Y^2 + 2lz}\right)^{{\textstyle \frac{D-3-k}{2}}}
\label{eq:dpot}\\
&\propto& \sqrt{l}
\int_0^1 \frac{dz \, z}{\sqrt{z} \, \sqrt{1-z}} 
G_{9-k}(Y^2 + 2 l z)
 \label{eq:dpotprop}
\end{eqnarray}
As in the Neumann case the potential is proportional to the mass of
the added open string state and to a $9-k$ dimensional Green's
function.  However, putting back the powers of $\alpha^\prime$, the
argument of the Green's function is now $Y^2/2\alpha^\prime + 2 l z$,
and there remains an integral over $z$ running from $0$ to $1$.  In
other words, the brane acts like an extended object with thickness
of order $\sqrt{4l\alpha^\prime}$ 
because of the remaining integral over $z$.
It can be easily checked that this effect is not an artifact of the
leading order in $q$ approximation.  The static potential computed
here does not depend on the relative orientation of the excitation and
the separation $Y$ of the branes, since we are only considering states
polarized orthogonal to $Y$. (See Section~\ref{sec:amp}.)  Indeed, the
form of the potential is what is expected for a body with
extent of order
$\sqrt{4l\alpha^\prime}$ transverse to the line of separation.
This is a very direct demonstration of how the
transversely polarized massive states endow D-branes with thickness.
Note that our study of the longitudinally polarized states did not
reveal this effect because such states lie entirely within
the world volume.

\section{Conclusion and Discussion}
\label{sec:impl}
In this paper we have computed the one-loop diagram for leading Regge
trajectory open string states attached to a D-brane.  Generically,
such states are expected to be quite unstable because they are far
from BPS saturation.  Nevertheless, we showed that there are large
kinematic suppressions which can partially stabilize such systems.
States that are polarized parallel to a brane have decay rates
suppressed by powers of logarithms of the energy.  Remarkably, states
polarized normal to a brane have leading decay rates that often
{\em decrease} with energy.  In particular, leading Regge trajectory
states on 0-branes cannot decay by splitting into two lighter states
and must rely on higher order processes to de-excite.

The splitting of strings discussed in this paper leads to
equipartition of energy amongst the degrees of freedom living on the
brane.  In view of the large kinematic suppressions found above, it
would be interesting to compute the rate for emission of closed
strings in order to see whether a highly excited RR soliton has a
chance to thermalize before a sizable portion of its energy is
radiated away.  It has been found that closed string emission from
D-branes with massless excitations is exponentially suppressed in the
energy of the emitted state~\cite{aki96a}.  This suggests that the
power law suppression of equipartition found in this paper is not
sufficient to impede thermalization.  It would be very interesting to
carry out similar analyses of the regular black hole configurations
in~\cite{strom96a}~-~\cite{vijay96b}.

The large suppressions found in this paper arise from restriction of
momenta to lie parallel the brane and mean that the world volume
decays of transversely polarized states may largely be ignored for
very heavy states.  The dominant decay mode for these states will be
into a single outgoing massless closed string.  This process may be
computed on a disk with one boundary and one bulk operator or by
looking at the closed string poles of the non-planar loop diagram.
Sub-leading Regge trajectory decays will suffer less severe
suppression.  This is because, as discussed in
Section~\ref{sec:trans}, some of the one-loop kinematic constraints
translate into angular momentum conservation in tree level diagrams.
These constraints are less severe for sub-leading trajectories.

In Section~\ref{sec:forces} we have computed the static force between
a brane excited by a leading Regge trajectory state and another in its
ground state.  The force is non-vanishing because the system is no
longer BPS saturated.  When the excited string is polarized normal to
the brane the force law shows evidence that the brane acts like an
extended body with size of order $\sqrt{\alpha^\prime}$.  These
excitations can also be expected to affect the tachyonic instability
that occurs when a brane and anti-brane approach each
other~\cite{susskind9511,green9604}.  It would be interesting to
understand how exciting a brane affects this singularity.

\section{Acknowledgments}
We would like to thank Aki Hashimoto for collaboration in the early
part of this project and Finn Larsen for fruitful discussions.
We are also grateful to Curt Callan, Eric Sharpe and Larus Thorlacius 
for helpful conversations.  V.B. was supported in part by DOE grant
DE-FG02-91ER40671. 
I.R.K. was supported in part by DOE grant DE-FG02-91ER40671, the NSF
Presidential Young Investigator Award PHY-9157482, and the James S.{}
McDonnell Foundation grant No.{} 91-48.


\end{document}